# Resonance Magnetoelectric Effects in Layered Magnetostrictive-Piezoelectric Composites

Ì. I. Bichurin, D. A. Filippov and V. M. Petrov

*Department of Physics and Engineering, Novgorod State University,*

*B. S. Peterburgskaya St. 41, 173003 Veliky Novgorod, Russia*

V. M. Laletin

*Institute of Technical Acoustics, National Academy of*

*Sciences of Belarus, 210717 Vitebsk, Belarus*

G. Srinivasan

*Physics Department, Oakland University Rochester, MI 48309, USA*

**ABSTRACT**

Magnetoelectric interactions in bilayers of magnetostrictive and piezoelectric phases are mediated by mechanical deformation.  Here we discuss the theory and companion data for magnetoelectric (ME) coupling at electromechanical resonance (EMR) in a ferrite-lead zirconate titanate (PZT) bilayer.  Estimated ME voltage coefficient versus frequency profiles for nickel, cobalt, or lithium ferrite and PZT reveal a giant ME effect at EMR with the highest coupling expected for cobalt ferrite-PZT. Measurements of resonance ME coupling have been carried out on layered and bulk composites of nickel ferrite-PZT.  We observe a factor of 40-600 increase in ME voltage coefficient at EMR compared to low frequency values.  Theoretical ME voltage coefficients versus frequency profiles are in excellent agreement with data. The resonance ME effect is therefore a novel tool for enhancing the field conversion efficiency in the composites.

Magnetoelectric materials facilitate the conversion between energies stored in magnetic and electric fields. The effect requires the presence of both long-range magnetic order and permanent dipole moment. Very few single-phase materials are magnetoelectric, but the effect is usually weak.[1] A much stronger magnetoelectric (ME) effect could be realized in a composite of magnetostrictive and piezoelectric phases in which the ME coupling is mediated by mechanical stress. Magnetostriction induced mechanical strain results in piezoelectric induced electric fields. The ME composites of interest in the past were bulk samples of ferrites with $BaTiO_3$ or lead zirconate titanate (PZT).[1,2] Bulk composites in general show ME coupling much smaller than predicted values due to leakage currents associated with low resistivity for ferrites. Problems inherent to bulk composites could easily be eliminated in a layered structure since high electrical resistivity leads to the absence of leakage current. Such structures are also easy to pole in an electric field, thereby strengthening the piezoelectric and ME effect. Our recent studies on bilayers and multilayers of ferrite-PZT and lanthanum manganite-PZT show evidence for ME coupling much stronger than in bulk samples.[3-5]

We propose here a unique and novel technique for further enhancing the field conversion efficiency of the composite, i.e., ME effects at electromechanical resonance (EMR). This work constitutes the first theoretical effort and companion first data for the phenomenon in ferrite-PZT layered samples. The resonance ME effect is similar in nature to the standard effect, i.e. an induced polarization under the action of an ac magnetic field. But the ac field here is tuned to the electromechanical resonance frequency. As the dynamic magnetostriction is responsible for the electromagnetic coupling, EMR leads to significant increasing in the ME voltage coefficients. The technique was implemented once in the past for bulk composites in order to distinguish weak ME signals from noise.[1]

In a ferrite-PZT composite, an applied ac magnetic field $\delta H$ produces dynamic deformation due to magnetostriction and results in an electric field $\delta E$ due to piezoelectric effect. The induced polarization $\delta P$ is related to the field $\delta H$ by $\delta P = \alpha \, \delta H$, where $\alpha$ is the ME-susceptibility. The parameter measured in most studies is the ME voltage coefficient $\alpha_E = \delta E/\delta H$ and is related to $\alpha$ by the expression $\alpha = \varepsilon_o \varepsilon_r \alpha_E$, where $\varepsilon_r$ is the relative permittivity of the material. There have been few reports on the theory of ME coupling in layered samples. We recently proposed a model for low frequency effects in ferrite-PZT bilayers.[6-8] An averaging method was used to estimate effective material parameters and ME voltage coefficients $\alpha_E$. It was shown that $\alpha_E$ depended on the piezoelectric, piezomagnetic and elastic constants for the two phases. The model predicted $\alpha_E$ on the order of 100-4000 mV/cm Oe and the strongest ME effect in cobalt ferrite-PZT among ferrite-based composites. These predictions were in agreement with observations in thick film multilayers of manganite-PZT and ferrite-PZT with $\alpha_E$ =30-1500 mV/cm Oe.[3-5]

The theory detailed here predicts giant magnetoelectric interactions in ferrite-PZT bilayers at frequencies corresponding to electromechanical resonance. The model is for radial modes in the bilayer. A longitudinal field configuration in which all the fields are parallel to each other and perpendicular to the sample plane is considered. An averaging procedure is employed to obtain the composite parameters and the longitudinal ME voltage coefficient $\alpha_{E,L}$. The theory is applied to bilayers containing ferrites of importance, i.e., cobalt ferrite (CFO) due to high magnetostriction, nickel ferrite (NFO) due to strong magnetomechanical coupling, and lithium ferrite (LFO) because of low loss characteristics. Based on our theory, one expects a resonance character in $\alpha_{E,L}$ versus frequency profile with a maximum $\alpha_{E,L}$ that is a factor of 40-70 higher than low frequency values. The strongest ME coupling is predicted for CFO-PZT and the weakest for LFO-PZT. We also discuss measurements on resonance ME coupling in both layered and bulk composites. Bulk composites are found to show a much stronger EMR-assisted enhancement in ME coupling compared to layered samples. The theoretical estimates are in excellent agreement with data for multilayers of NFO-PZT. Details on the theory and data are provided here.

We assume the composite to be a homogeneous medium that can be described by effective parameters, such as compliance, piezoelectric and magnetostrictive coefficients that are determined from parameters for the magnetostrictive and piezoelectric phases. The assumption is valid when the layer thickness is small compared to wavelengths for the acoustic modes and is certainly true for electromechanical resonance that occurs at 100-500 kHz. We consider a ferrite–PZT bilayer in the form of thin disk of radius *R* and thickness *d*. The electrodes on the bilayer are assumed to be of negligible thickness. The sample is poled and magnetized perpendicular to the sample plane and the ac magnetic field is applied parallel to poling direction ($z$). The ac magnetic field induces harmonic waves in the radial or thickness modes. It is supposed that $d \ll R$ so that only radial modes are considered. For a thin disk, it is possible to neglect any pressure variation along the $z$-axis. The axial symmetry results in nonzero components of the pressure and strain tensors $T_{rr}$, $T_{èè}$, $S_{rr}$ and $S_{èè}$. The generalized Hooke's law and corresponding equations have the following form:

$$S_{rr} = s_{11}T_{rr} + s_{12}T_{qq} + d_{31}E_z + q_{31}H_z,$$

$$S_{qq} = s_{12}T_{rr} + s_{11}T_{qq} + d_{31}E_z + q_{31}H_z, \quad (1)$$

$$D_z = d_{31}(T_{rr} + T_{qq}) + e_{33}E_z + a_{33}H_z,$$

where $S_{rr} = \dfrac{\partial u_r}{\partial r}$, $S_{qq} = \dfrac{1}{r}u_r$, $u_i$ is displacement coordinate of medium, $T_{ij}$ is stress tensor component, $s_{ij}$ is compliance coefficient, $q_{ij}$ and $d_{ij}$ are piezomagnetic and piezoelectric coefficients, $D_i$ is electric displacement component, $e_{ij}$ is permittivity matrix, and $á_{33}$ is the ME coefficient..

The equation of elastodynamics has the following form for the radial propagating mode:

$$\frac{dT_{rr}}{dr} + \frac{1}{r}(T_{rr} - T_{qq}) + rw^2 u_r = 0, \quad (2)$$

where $r$ is the density and $w$ is the angular frequency. Solutions of the Eq. (2), taking into account Eq. (1), can be presented as superposition of the first and second order Bessel functions.

Using the boundary conditions: $u_r = 0$ at $r = 0$ and $T_{rr} = 0$ at $r = R$ we get expressions for $T_{rr}$ and $T_{èè}$

$$T_{rr} = \frac{1}{s_{11}(1-n)}\left[\frac{kRJ_0(kr) - (1-n)\dfrac{R}{r}J_1(kr)}{\Delta_r} - 1\right] \cdot (q_{31}H_z + d_{31}E_z) \quad (3)$$

$$T_{qq} = \frac{1}{s_{11}(1-\nu)}\left[\frac{\nu k R J_0(kr) + (1-\nu)\frac{R}{r}J_1(kr)}{\Delta_r} - 1\right]\cdot(q_{31}H_z + d_{31}E_z), \quad (4)$$

where $k = \sqrt{\rho s_{11}(1-\nu^2)}\omega$, $\nu = -s_{12}/s_{11}$ is the Poisson's ratio, and $\Delta_r = kRJ_0(kR) - (1-\nu)J_1(kR)$.

The electric field, obtained by taking into consideration open circuit conditions, i.e. $\int_S D_n dS = 0$, where $S$ is electrode plane, is

$$E_z = -\frac{1}{|\Delta_a|}\left[\frac{2d_{31}q_{31}}{e_{33}s_{11}(1-\nu)}\left(\frac{(1+\nu)J_1(kR)}{\Delta_r}-1\right) + \frac{\alpha_{33}}{e_{33}}\right]H_z, \quad (5)$$

where $\Delta_a = 1 - K_p^2 + K_p^2(1+\nu)J_1(kR)/\Delta_r + i\Gamma$, (6)

$K_p^2 = \frac{2d_{31}^2}{e_{33}s_{11}(1-\nu)}$ is the coefficient of electromechanical coupling for radial mode and $\tilde{A}$ is the loss factor. Finally, the longitudinal ME voltage coefficient $\alpha_{E,L} = dE_z/dH_z$, is obtained as

$$\alpha_{E,L} = -\frac{1}{|\Delta_a|}\left[\frac{2d_{31}q_{31}}{e_{33}s_{11}(1-\nu)}\left(\frac{(1+\nu)J_1(kR)}{\Delta_r}-1\right) + \frac{\alpha_{33}}{e_{33}}\right]. \quad (7)$$

The frequency dependence of $\alpha_{E,L}$ has a resonance character. The resonance frequency is determined by the condition $\text{Re}\,\Delta_a = 0$ and the ME voltage coefficient is expected to show a peak at this frequency. It is the so-called anti-resonance condition. It follows from Eq. (6) that the resonance frequency depends on sample radius and the material parameters: compliances $s_{11}$ and $s_{12}$, density $\tilde{n}$ and coefficient of electromechanical coupling for radial mode $K_p$. The peak value of $\alpha_{E,L}$ and resonance line width are determined by effective piezomagnetic ($q_{31}$) and piezoelectric($d_{31}$) coefficients, compliances, permittivity and loss factor.

We next apply the theory to representative ferrite-PZT bilayers. The ferrites considered include cobalt ferrite because of high magnetostriction and piezomagnetic coupling,[1] nickel ferrite due to strong magneto-mechanical coupling,[4,5] and lithium ferrite. Estimated $\alpha_{E,L}$ versus frequency are shown in Fig.1 for parameters in Table 1. Calculations are for PZT volume fraction of 0.7 and effective composite parameters in Ref.6. We assumed a bias field H that corresponds to maximum piezomagnetic coupling q. The sample diameter has been chosen in such a way so as to separate the EMR for the three bilayers. A resonance character is clearly evident in Fig.1; the resonance $\alpha_{E,L}$ is the highest for CFO-PZT due to high piezomagnetic coupling and is the lowest for LFO-PZT. The width at half-maximum range from 8 to 12 kHz and the quality-factor for the resonance is in the range 25-40. The most important inference from Fig.1 is the overall enhancement in $\alpha_{E,L}$ at resonance, a factor of 40-70 higher than off-resonance values.

Next we consider companion data on resonance ME effects at EMR. We provide here only results relevant to the theory. Exhaustive details on sample synthesis and low frequency characterization have been discussed in Ref.3-5. Detailed high frequency characterization will be published elsewhere.[9] Investigations were carried out on multilayer composites of NFO-PZT. A 10 mm diameter sample that contained 11 layers of 13 μm thick NFO and 10 layers of PZT with a thickness of 26 ìm was used. It was poled with an electric field *E* perpendicular to the sample plane. A bias magnetic field *H* and an ac field *dH* were applied perpendicular to the sample plane. The induced electric field *dE* was measured across the sample thickness. We first measured $a_{E,L}$ vs *H* profile for ac fields at 100 Hz and the results are shown in Fig.2. One observes a rapid in increase in $a_{E,L}$ with increasing *H* and is followed by a decrease in $a_{E,L}$ to zero value. As discussed in Refs.3-5, $a_{E,L}$ essentially tracks the variation of *q* with *H* and it vanishes when the magnetostriction attains saturation. Notice the maximum in $a_{E,L}$ for a certain $H=H_m$. Then for *H* set at $H_m$, we measured $a_{E,L}$ as a function of the frequency (f = 0–500 kHz) of the ac field δH. Figure 2 shows such data for the NFO-PZT multilayer. Theoretical estimates based on the current theory are shown for comparison. The resonance frequency is 350 kHz, in agreement with calculated value for the sample dimension and composite parameters. The theoretical resonance profile for a loss factor $\tilde{A} = 0.08$ tracks the observed variation in $\alpha_{E,L}$ with f. The most significant inference in Fig.2 is the realization of predicted giant ME interactions in the composite at EMR. The $\alpha_{E,L}$ value at resonance is 1200 mV/cm Oe and must be compared with the low-frequency value of 30 mV/cm Oe. Although the estimated $\alpha_{E,L}$ at EMR is 15% higher than the measured value, the overall agreement between theory and data is excellent.

Another significant finding of relevance to the present study concerns the resonance ME effect in bulk ferrite-PZT composites. We performed such studies on a representative sample consisting of modified nickel ferrite and PZT. It was necessary to modify NFO with a combination of substitutions and Fe deficiency to increase the electrical resistivity. The high resistivity leads to excellent poling characteristics and improves ME interactions.[10] We prepared samples with PZT concentration varying from 10 to 90 wt.%. Magnetoelectric voltage coefficients were measured at low frequencies (1 kHz) and at EMR. The peak $\alpha_{E,L}$ values as a function of PZT concentration $w$ are shown in Fig.3. The low frequency data indicate the absence of ME effects in pure ferrite or PZT and a concentration independent *$a_{E,L}$* for $w$=40-80%. The bulk samples show *$a_{E,L}$* values that are comparable to results in Fig.2(a) for layered NFO-PZT. Upon increasing the frequency of ac field to EMR, we infer from Fig.3 a dramatic strengthening of ME interaction; *$a_{E,L}$* increases by a factor of 600 at resonance. The ME voltage coefficient at resonance is as high as 23000 mV/cm Oe. The overall enhancement in *$a_{E,L}$* at resonance is much higher in bulk samples than for layered systems (Fig.2). The ME coefficients for layered and bulk samples of NFO-PZT in Figs.2 and 3 are one of the largest ever reported for any composites. In some recent studies, ferrites have been replaced by terfenol, a highly magnetostrictive and piezomagnetic alloy. Layered terfenol-PZT samples show off-resonance *$a_{E,L}$* on the order of 4680 mV/cm Oe, a value comparable to resonance values in ferrite-PZT.[11]

In conclusion, a theoretical model has been developed for ME effects in layered composites at electromechanical resonance. The theoretically predicted giant ME interactions at resonance is in agreement with the data for nickel ferrite-PZT samples.

This work is supported by grants from the Russian Ministry of Education (Å02-3.4-278), the Universities of Russia Foundation (UNR 01.01.007) and the National Science Foundation (DMR-0322254).

Table 1: Compliance coefficient *s*, piezomagnetic coupling *q*, piezoelectric coefficient *d* and petmittivity *e* for cobalt ferrite (CFO), nickel ferrite (NFO), lithium ferrite (LFO) and lead zirconate titanate (PZT).

| Material | $s_{11}$ ($10^{-12}$ m$^2$/N) | $s_{12}$ ($10^{-12}$ m$^2$/N) | $q_{31}$ ($10^{-12}$ m/A) | $d_{31}$ ($10^{-12}$ m/V) | $\varepsilon_{33}/\varepsilon_0$ |
|---|---|---|---|---|---|
| CFO | 6.5 | -2.4 | 556 | | 10 |
| NFO | 6.5 | -2.4 | 125 | | 10 |
| LFO | 3.3 | -1.65 | -12.5 | | 10 |
| PZT | 15.3 | -5 | | -175 | 1750 |

**Caption for Figures**

Fig.1: Theoretical longitudinal magnetoelectric (ME) voltage coefficient $a_{E,L} = dE/dH$ as a function of the frequency of the ac field $dH$ applied to bilayers of ferrite and lead zirconate titanate (PZT). The poling electric field is perpendicular to the bilayer plane. The bias field $H$ and the resulting ac electric field are also along the same direction. Estimates are for sample thickness much smaller than the radius $R$. The radius of the bilayer samples are 6 mm for cobalt ferrite (CFO)-PZT, 5 mm for nickel ferrite (NFO)-PZT and 4 mm for lithium ferrite (LFO)-PZT. Notice the resonance-like behavior for $a_{E,L}$ vs $f$ with the peak value at the electromechanical resonance (EMR) frequency. Parameters used for the estimates are given in Table 1.

Fig.2: (a) Data from Ref.3 for low frequency (100 Hz) $a_{E,L}$ versus the bias field $H$ for a multilayer sample of NFO-PZT. The dashed line is guide to the eye. (b) Data on the frequency dependence of $\alpha_{E,L}$ for the multilayer sample. The sample contained 11 layers of ferrite of thickness 13 μm and 10 layers of PZT with a thickness of 26 μm. The bias field H was set at 1050 Oe, the field at which low frequency $\alpha_{E,L}$ is maximum. The solid line is the calculated values based on the present theory for a bilayer (Fig.1).

Fig. 3: Data on off-resonance and resonance ME effects in bulk samples of modified nickel ferrite-PZT. Data for samples with a series of PZT amount $w$=0-100 wt.% are shown. Peak values of $a_{E,L}$ at low frequencies and at EMR were measured from data as in Fig.2. The lines are guide to the eye.

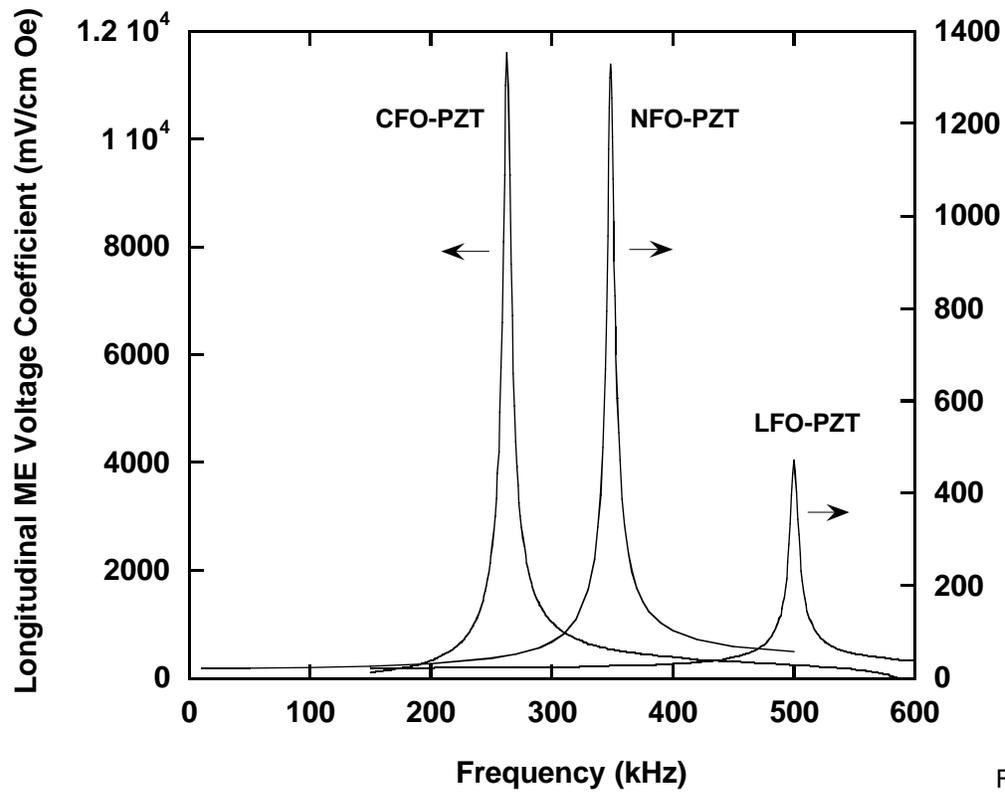

Fig. 1

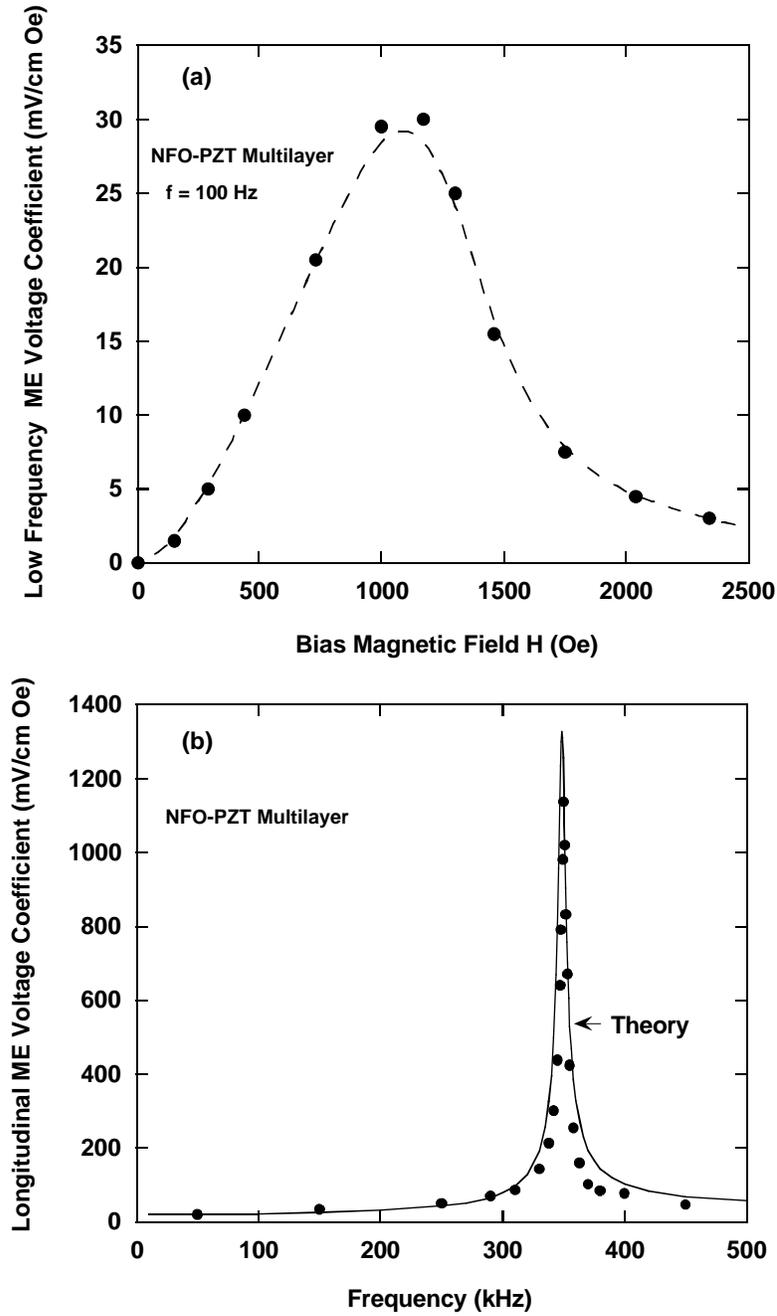

Fig. 2

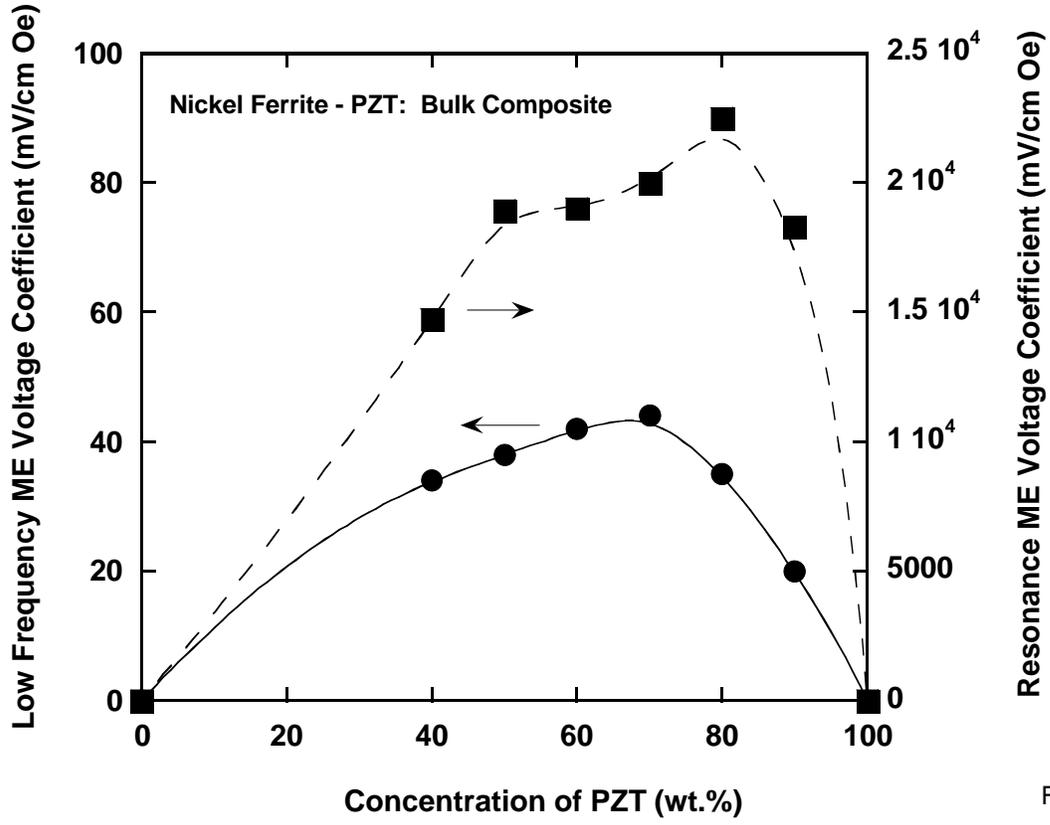

Fig. 3